\documentclass[]{aa}
\pdfoutput=1
\usepackage[version=3]{mhchem}

\usepackage{graphicx}
\usepackage{txfonts}
\usepackage[]{natbib}

\newcommand{\pccp}{PCCP}%

\begin{document}
\title{On the relevance of the H$_2$ + O reaction pathway for the surface formation of interstellar water}
\subtitle{A combined experimental and modeling study}
\titlerunning{the H$_2$ + O reaction pathway for solid interstellar water}
\author{T. Lamberts\inst{1,2}, H. M. Cuppen\inst{2}, G. Fedoseev\inst{1}, S. Ioppolo\inst{2, 3}, K.-J. Chuang\inst{1}, H. Linnartz\inst{1}}
\authorrunning{Lamberts et al.}
\institute{Raymond and Beverly Sackler Laboratory for Astrophysics, Leiden Observatory, University of Leiden, P.O. Box 9513, NL 2300 RA Leiden, The Netherlands. \email{lamberts@strw.leidenuniv.nl} \and Faculty of Science, Radboud University Nijmegen, IMM, P.O. Box 9010, NL 6500 GL Nijmegen, The Netherlands
\and Division of Geological and Planetary Sciences, California Institute of Technology, 1200 E. California Blvd., Pasadena, California 91125, USA}
\date{}

\abstract{The formation of interstellar water has been commonly accepted to occur on the surfaces of icy dust grains in dark molecular clouds at low temperatures (10-20~K) , involving hydrogenation reactions of oxygen allotropes. As a result of the large abundances of molecular hydrogen and atomic oxygen in these regions, the reaction \ce{H2 + O} has been proposed to contribute significantly to the formation of water as well. However, gas phase experiments and calculations, as well as solid-phase experimental work contradict this hypothesis. Here, we use precisely executed temperature programmed desorption (TPD) experiments in an ultra-high vacuum setup combined with kinetic Monte Carlo simulations to establish an upper limit of the water production starting from \ce{H2} and \ce{O}. These reactants are brought together in a matrix of \ce{CO2} in a series of (control) experiments at different temperatures and with different isotopological compositions. The amount of water detected with the quadrupole mass spectrometer upon TPD is found to originate mainly from contamination in the chamber itself. However, if water is produced in small quantities on the surface through \ce{H2 + O}, this can only be explained by a combined classical and tunneled reaction mechanism. An absolutely conservative upper limit for the reaction rate is derived with a microscopic kinetic Monte Carlo model that converts the upper limit into a maximal possible reaction rate. Incorporating this rate into simulations run for astrochemically relevant parameters, shows that the upper limit to the contribution of the reaction \ce{H2 + O} in OH, and hence water formation, is 11\% in dense interstellar clouds. Our combined experimental and theoretical results indicate however, that this contribution is likely to be much lower.}
  

\maketitle

\section{Introduction}
The formation of interstellar water is commonly believed to occur mostly on the surfaces of icy dust grains in dark molecular clouds where the temperatures range typically between 10 and 20~K. In recent years, several studies have been focusing on the reaction of atomic hydrogen with O, \ce{O2}, and \ce{O3} in interstellar ice analogues, both experimentally and through surface models \citep{Hiraoka:1998, Dulieu:2010, Miyauchi:2008, Ioppolo:2008, Oba:2009, Ioppolo:2010, Cuppen:2010B, Mokrane:2009, Romanzin:2011, Oba:2012, Lamberts:2013}. A possibly interesting alternative pathway to form water under interstellar conditions starts from the reaction 
\begin{equation}
\ce{H2} + \ce{O} \rightarrow \ce{OH} + \ce{H}\;\; \tag{R1}\label{R1}
\end{equation}
and is followed by 
\begin{equation}
\ce{OH} + \ce{H} \rightarrow \ce{H2O} \;\; \tag{R2} \;  \label{R2}
\end{equation}
or
\begin{equation}
\ce{OH} + \ce{H2} \rightarrow \ce{H2O} + \ce{H}. \;\; \tag{R3} 
\end{equation}
Reaction R1 has been proposed to contribute significantly to the formation of water since molecular hydrogen and atomic oxygen are both abundantly present in the dense regions of the interstellar medium \citep{Cazaux:2010, Cazaux:2011}. Additionally, \citet{Cazaux:2010} proposed this reaction to be important for deuterium enrichment during water formation. Conceptually the interaction between \ce{H2} and the surface could aid in breaking the H-H bond. The reaction is, however, endothermic by 960 K, making it intuitively unlikely to occur in the low temperature regime. Moreover, a theoretical barrier in the gas phase of approximately 7000 K is predicted for the case that both O and \ce{H2} are in the ground state \citep{Rogers:2000}. Gas-phase experimental work also predicts high barriers ($\sim$3000 K) as reviewed by \cite{Baulch:1992}. Barriers of this order of magnitude lead to thermally induced reaction rates that are so slow that their contribution to the full chemical reaction network becomes negligible even over the long interstellar time scales of several million years \citep{Bergin:2007}. It should be noted that at low temperatures tunneling may play an important role, but tunneling through the barrier of an endothermic reaction can only take place if the reactants have an initial energy equal to or larger than the endothermicity \citep{Arnaut:2006}.

For the reasons mentioned above, reaction R1 was excluded in the reaction scheme used by \citet{Cuppen:2007A} who studied the formation of ice mantles on interstellar grains. Recent solid state laboratory studies by \citet{Oba:2012} showed no detectable production of \ce{H2O} by means of infrared spectroscopy upon co-deposition of \ce{H2} and O atoms, which motivated \citet{Taquet:2013} to exclude it from their ice chemistry reaction network as well.

Here ultra-high vacuum (UHV) surface chemistry experiments are carried out at low temperature in conjunction with kinetic Monte Carlo modeling to clarify the ambiguity in the importance of the reaction \ce{H2} + O under interstellar conditions. 

\section{Calculation of the reaction rate}\label{CalculateRate}
\begin{figure}[ht]
\centering
 \includegraphics[width=0.45\textwidth]{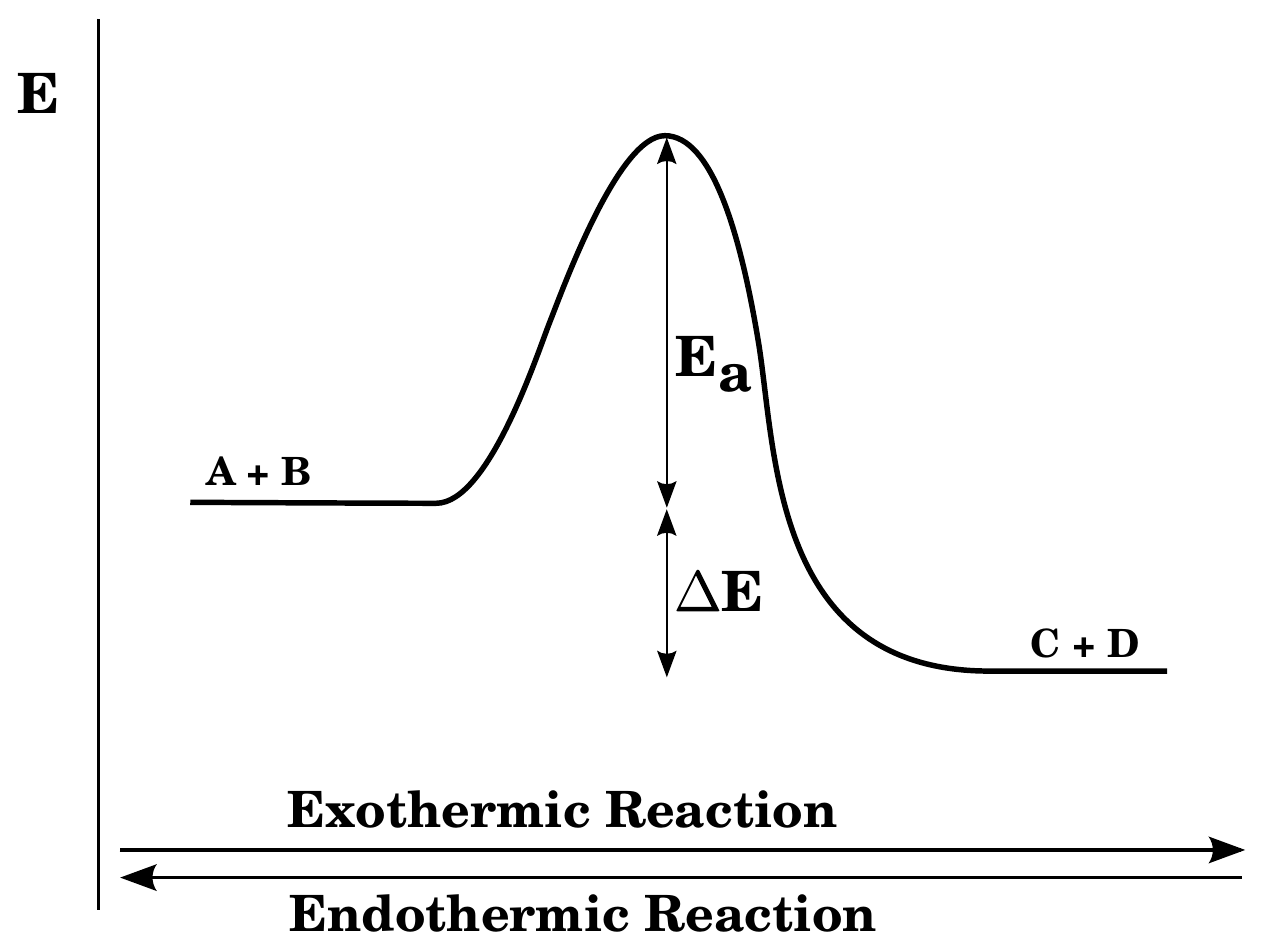}
\caption{Schematic representation of the energy level diagram of an exothermic and endothermic reaction. }\label{ReactionPathway}
\end{figure}

Reactions are often considered to take place along pathways such as those shown in Fig.~\ref{ReactionPathway}. The reaction coordinate is depicted on the horizontal axis, energy on the vertical axis, $\Delta E$ indicates the difference in potential energy between reactants (A + B) and products (C + D) and the reaction rate is determined by the barrier or activation energy, $E_{\text{a}}$.
In astrochemical models it is common to use a straightforward expression for the calculation of a reaction rate as a result of the large chemical networks involved \citep{Garrod:2006A}. The calculation of the reaction rates therefore often involves making a rather arbitrary choice between the expression for classically (\emph{i.e.}, thermally) activated reactions
\begin{equation}
k_{\text{therm.}} = \nu\exp\left( - \frac{E_{\text{a}}}{T} \right) \label{thermal}
\end{equation}
and the expression for tunneling of a free particle through a rectangular barrier \citep{Bell:1980}
\begin{equation}
k_{\text{exo. tunn.}} = \nu\exp\left( - \frac{2\,a}{\hbar} \sqrt{2\,\mu\,E_{\text{a}}} \right). \label{tunneled}
\end{equation}
Typically the trial frequency $\nu$ is approximated by the standard value for physisorbed species, ${kT}/{h} \approx 10^{12}$ s$^{-1}$ and a barrier width $a$ of 1 \AA~is chosen.
In the expression for the tunneling rate the reduced mass, $\mu_{\text{red}}$, is usually taken to be the reduced mass of the total reacting system without taking into account the mutual orientation of the reactants. The mass should, however, be affiliated with the reaction coordinate involved as was done in recent work of a linear bimolecular atom-transfer reaction leading to an effective mass, $\mu_{\text{eff}}$ \citep{Oba:2012}. In the case of reaction R1 the difference between the reduced and the effective mass gives rise to a substantial increase of the reaction rate (see also Table~\ref{rates}).

Tunneling rates for endothermic reactions, $k_{\text{endo. tunn.}}$ (see Fig.~\ref{ReactionPathway}), need to be calculated as a combination of both Eqs.~\ref{thermal} and \ref{tunneled}, where the classical contribution accounts for the part of the reaction barrier which lies below the endothermicity and the tunneled contribution for that above \citep{Arnaut:2006}. This can be derived from arguments of detailed balance (or microscopic reversibility): in equilibrium the net flux between every pair of states is zero. The reaction rates should then obey the condition
\begin{equation}
\frac{k_{\text{endo. tunn.}}}{k_{\text{exo. tunn.}}} = \exp\left(- \frac{\Delta E}{T} \right) \;.
\end{equation}
and hence, following the definition for $E_{\text{a}}$ from Fig.~\ref{ReactionPathway},
\begin{align}
k_{\text{endo. tunn.}} &= k_{\text{exo. tunn.}} \exp\left(- \frac{\Delta E}{T} \right) \nonumber \\
&= \nu\exp\left( - \frac{2\,a}{\hbar} \sqrt{2\,\mu\,E_{\text{a}}} \right) \exp\left(- \frac{\Delta E}{T} \right) \;.\label{endotunn}
\end{align}
The comparison between these various ways of calculating the reaction rate spans a large range as outlined in Table~\ref{rates}.  
A more accurate way to calculate reaction rates also takes into account the shape of the barrier, examples of which are the usage of the Eckart model by \citet{Taquet:2013} or the implementation of instanton theory by \citet{Andersson:2011}. This results in modified tunneling reaction rates with differences up to several orders of magnitude. Depending on the expression used, the resulting reaction rate can be substantially different. The ambiguity makes it hard to interpret these values in terms of their astronomical relevance. One way to partially circumvent this is to make use of upper (or lower) limits, determined experimentally. 

In the following sections we will use laboratory experiments combined with microscopic kinetic Monte Carlo simulations to constrain the reaction rate of reaction R1. Subsequently the resulting reaction rate is incorporated into the same kinetic Monte Carlo model, but run with physical parameters relevant to the interstellar medium to test its astronomical significance.

\begin{table}[ht]
 \centering\small
\caption{Calculated reaction rates for the reaction \ce{H2 + O} assuming classical and tunneled contributions.}\label{rates}
\begin{tabular}{llllll}
\hline
Eqn.~used		& $T$		 & $\Delta E$		& $E_\text{a}$& $\mu$			& Rate	\\
			& (K)		& (K)	& 	(K)	& 		& (s$^{-1}$) \\
\hline
(\ref{thermal})\; $k_{\text{therm.}}$ 	& 10		& 960	& (...)	& (...)                	& $5.2\times 10^{-119}$ \\
(\ref{thermal})\; $k_{\text{therm.}}$ 	& 13.5 		& 960	& (...)	& (...)                	& $3.1\times 10^{-85}$  \\
(\ref{endotunn})\; $k_{\text{endo. tunn.}}$& 10 	& 960	& 2040$^{a}$	& $\mu_{\text{eff}}=0.47$ & $5.0\times 10^{-36}$ \\
(\ref{endotunn})\; $k_{\text{endo. tunn.}}$& 13.5  	& 960	& 2040$^{a}$	& $\mu_{\text{eff}}=0.47$	& $3.2\times 10^{-25}$ \\
(\ref{tunneled})\; $k_{\text{exo. tunn.}}$ & (...)	& (...)	& 3000 (1) & $\mu_{\text{red}}=1.78$	               & $1.2\times 10^{-1}$ \\
(\ref{tunneled})\; $k_{\text{exo. tunn.}}$& (...) 	& (...)	& 3000 (1)  & $\mu_{\text{eff}}=0.47$              		& $2.3\times 10^{+5}$ \\
\hline
\end{tabular}
\tablefoot{
\tablefoottext{a}{The total barrier of the reaction is the combination of the endothermicity, $\Delta E$, and the barrier itself, $E_\text{a}$, which amounts in total to $\sim 3000$~K (1). 
}}
\tablebib{(1)~ \citet{Baulch:1992}}
\end{table}

\section{Experiments}
\subsection{Methods}
Experiments are performed using the SURFRESIDE$^2$ setup which allows for the systematic investigation of solid state reactions leading to the formation of molecules of astrophysical interest at cryogenic temperatures. SURFRESIDE$^2$ consists of three UHV chambers with a room-temperature base-pressure between $10^{-9}-10^{-10}$ mbar. The setup has already been extensively described in \citet{Ioppolo:2013} and therefore only a brief description of the used procedure is given here. 
A rotatable gold-coated copper substrate in the center of the main chamber is cooled to 13.5-14.0 K using a He closed-cycle cryostat with an absolute temperature accuracy of $\leq$ 2~K. This temperature is around the lower limit of what can be reached under our experimental conditions and is chosen to minimize the diffusion of the oxygen atoms, but simultaneously have a high lifetime of \ce{H2} and O on the surface.
To study the solid state reaction pathway \ce{H2 + O}, the reactants need to be deposited on a surface, while simultaneously preventing the competing reactions \ce{O + O -> O2} and \ce{O + O2 -> O3}. This is achieved by using a matrix consisting of \ce{CO2} molecules and an overabundance of molecular hydrogen. A full experiment starts with the preparation of all selected gases in separate pre-pumped ($\leq 10^{-5}$ mbar) dosing lines. Subsequently a co-deposition of \ce{H2}, O and \ce{CO2} is performed. Room temperature carbon dioxide (Praxair 99.996\%) is deposited through a metal deposition line under an angle of 90$^{\circ}$.
Room-temperature molecular hydrogen (Praxair 99.999\%) is deposited on the surface through an UHV beam line with an angle of 45$^{\circ}$ with respect to the surface. Oxygen atoms are generated from \ce{^{18}O2} (Aldrich 99\%) in another UHV beam line in a microwave plasma atom source (Oxford Scientific Ltd, see \citep{Anton:2000}) with an angle of 135$^{\circ}$ with respect to the surface. A custom made nose-shaped quartz-pipe is placed in between the atom sources and the substrate. The pipe is designed in such a way that all chemically active species that are in their electronic and/or ro-vibrationally excited states are quenched to room temperature before being deposited to the surface. Besides \ce{^{18}O} atoms also a (large) fraction of non-dissociated \ce{^{18}O2} is present in the beam. The UHV beam lines can be operated independently and are separated from the main chamber by metal shutters. All experiments and the corresponding atomic and molecular fluxes are listed in Table~\ref{allexps}. The effective O flux determination by \citet{Ioppolo:2013} was repeated and found to be reproducible: $2\times 10^{11}$ at cm$^{-2}$ s$^{-1}$ (uncertainty $\sim$30\%). Each (control) experiment is performed for a duration of 75 minutes. Experiments 1, 2 and 3 have been performed twice to check their reproducibility. The aim of these experiments is to determine an upper limit for the production of water during co-deposition.

\begin{table*}[ht]
\centering\small
\caption{List of (control) experiments and integrated baseline corrected QMS signals for m/z = 20 and 22,  \emph{i.e.} \ce{H2^{18}O} and \ce{D2^{18}O}, and the calculated \ce{H2^{18}O} abundance in ML.}
\label{allexps}
 \begin{tabular}{c|cclllc@{ }l|lll}
  \hline
{Nr.}	& \multicolumn{7}{c}{{Experimental parameters}} & \multicolumn{3}{|c}{{Integrated QMS signal}} \\ 
        & T & Time & \ce{CO2} flux          & \ce{^{18}O_2} flux     & \ce{^{18}O} flux               & \ce{H2}/\ce{D2} &  flux     & 70-105 	& 125-175 & \ce{H2^{18}O}	\\
        & (K) & (min) &  (mol cm$^{-2}$ s$^{-1}$)    & (mol cm$^{-2}$ s$^{-1}$)      &  (at cm$^{-2}$ s$^{-1}$)      		&  	&  (mol cm$^{-2}$ s$^{-1}$)  & (K)	& (K)  & (ML) \\
\hline
1$^a$ & 14 & 75 & $1.6\times 10^{14}$ &   $1.7\times 10^{12}$   &  $2\times 10^{11}$    & \ce{H2}	&  $2.2\times 10^{14}$  &   $2.1\times 10^{-9}$	& $2.1\times 10^{-9}$ 	& 0.26  \\
2$^b$ & 14 & 75 & $1.6\times 10^{14}$  &   $1.7\times 10^{12}$      & $2\times 10^{11}$  & -- 		& --            	&  $9.6\times 10^{-10}$	&  $6.4\times 10^{-10}$	& 0.10 \\
3$^b$ & 14 & 75 & --     		     &    --  		& --                    & \ce{H2}	& $2.2\times 10^{14}$   & $2.2\times 10^{-10}$ 	& $4.2\times 10^{-10}$ 	&  0.03     \\
4$^b$ & 14 & 75 & --            	     &       --              & --              & --\;		& --           	      	& $\sim0$ 		& $1.3\times 10^{-10}$	&  $\sim0$   \\
5$^a$ & 14 & 300 & $1.6\times 10^{14}$ &   $1.7\times 10^{12}$   &  $2\times 10^{11}$    & \ce{H2}	& $2.2\times 10^{14}$  	&  --$^c$	& --$^c$ 	& $\sim 1$ $^d$  \\
{6$^a$} & {17} & {75} & $1.6\times 10^{14}$  &  $1.7\times 10^{12}$   & $2\times 10^{11}$   & {\ce{H2}}	& $2.2\times 10^{14}$	&  $1.2\times 10^{-9}$ 	& $1.3\times 10^{-9}$	&  {0.15}      \\
7$^a$ & 35 & 75 & $1.6\times 10^{14}$  &   $1.7\times 10^{12}$   &  $2\times 10^{11}$   & \ce{H2}	& $2.2\times 10^{14}$	& $1.7\times 10^{-9}$ 	& $8.4\times 10^{-10}$	&  0.16     \\
8$^a$ & 50 & 75 & $1.6\times 10^{14}$  &   $1.7\times 10^{12}$   &  $2\times 10^{11}$   & \ce{H2}	& $2.2\times 10^{14}$	& $1.6\times 10^{-9}$ 	& $4.2\times 10^{-10}$	&  0.12     \\
9$^a$ & 14 & 75 & $1.6\times 10^{14}$  &   $1.7\times 10^{12}$   &  $2\times 10^{11}$   & \ce{D2}	& $1.2\times 10^{14}$	& -- $^e$ 	& $5.0\times 10^{-11}$ $^f$	&  --     \\
  &  &   &   &     &   & & & $1.8\times 10^{-9}$ $^g$ 	& $9.4\times 10^{-10}$ $^g$	&  0.17     \\
10$^a$ & 14 & 75 & $1.6\times 10^{14}$  &   $1.7\times 10^{12}$   &  $2\times 10^{11}$   & \ce{D2}	& $2.2\times 10^{14}$	& -- $^e$ 	& $1.5\times 10^{-10}$ $^f$	&  --     \\
  &  &   &   &     &   & & & $2.7\times 10^{-9}$ $^g$ 	& $9.4\times 10^{-10}$ $^g$	&  0.22     \\
\hline
\end{tabular}
\tablefoot{
The baseline corrected QMS signals are integrated over two different temperature ranges covering the \ce{CO2} and the \ce{H2O} desorption peaks, 70-105~K and 125-175~K. \\
\tablefoottext{a}{Experiment} 
\tablefoottext{b}{Control experiment} 	
\tablefoottext{c}{Different ramp rate} 	
\tablefoottext{d}{from RAIRS}
\tablefoottext{e}{ Overlaps with doubly ionized \ce{CO2}}
\tablefoottext{f}{ {m/z = 22}}
\tablefoottext{g}{ {m/z = 20}} }
\end{table*}	 	

SURFRESIDE$^2$ has two main analytical tools: (i) the ice composition is monitored \emph{in situ} by means of reflection absorption infrared spectroscopy (RAIRS) in the range between 4000 and 700 cm$^{-1}$ with a spectral resolution of 1 cm$^{-1}$; (ii) the main chamber gas-phase composition is monitored by a quadrupole mass spectrometer (QMS), which is placed behind the rotatable substrate. Here, we deposit a total of 0.9~ML O atoms per experiment, meaning that RAIRS can only be used if the reaction is indeed as efficient as claimed by the exothermic tunneled rate. RAIR difference spectra with respect to the bare substrate are recorded every 5 minutes, averaging over 512 scans. After the co-deposition is finished, the sample is rotated to face the QMS and a temperature programmed desorption (TPD) experiment at 1~K min$^{-1}$ is performed to monitor the desorption of the ice constituents. The QMS is typically used for the study of species that fall below the detection limit of RAIRS, \emph{i.e.}, sub-monolayer experiments. 

To convert the integrated area of the current (pressure) read by the QMS to a number of molecules desorbing from the sample, we performed several calibration experiments. 
Firstly, to relate the ice thickness to a QMS signal, we deposited layers of water of three different thicknesses at 13.5-14~K, followed by a TPD at the usual ramp of 1~K min$^{-1}$. Subsequently the water RAIRS signal at 3280 and 1660 cm$^{-1}$ of these three experiments is converted into a number of monolayers using the IR bandstrength. This is, however, not trivial due to the reflection mode of the IR spectrometer, which is setup dependent. The bandstrength of \ce{CO2} in reflection mode was determined through an isothermal desorption experiment by Ioppolo et al. (2013). A similar calibration experiment cannot be easily performed for \ce{H2O}, due to rearrangement of hydrogen bonds at high temperatures, changing the desorption profile. Therefore, the ratio between the transmission bandstrengths of \ce{CO2} and \ce{H2O} has been taken from Gerakines et al. (1995) to derive the bandstrengths in reflection mode for the 3280 and 1660 cm$^{-1}$ bands of water. Finally, the value for the integrated QMS signal, corresponding to one monolayer of desorbing water molecules, is determined by averaging over the three deposited water layers. 

The experiments are analysed by first performing a linear baseline correction between 115 and 195~K. Then, the mass 20 amu signal is integrated over two ranges; one centred at the \ce{CO2} desorption ($\sim$80 K) and one at the \ce{H2O} desorption ($\sim$140~K) given in Table~\ref{allexps}. The combined signal is converted to a number of produced monolayers, averaged over the two experiments performed and given in the last column of Table~\ref{allexps}.

In previous experiments not listed in Table~\ref{allexps}, we have used a different \ce{CO2} flux and another source of atomic oxygen, \ce{N2^{16}O}. The latter has as main advantage that the competing ozone channel is less likely to occur since there are no large amounts of \ce{O2} present in the plasma source. It does yield regular water (\ce{H2^{16}O}) which is hard to distinguish from the contamination present in all parts of the experimental setup. The use of \ce{^{18}O2} as a precursor of atomic oxygen would lead to the formation of \ce{H2^{18}O} that can be better distinguished from background water contamination. However, as previously mentioned, the resulting O-atom beam will have an over abundance of undissociated \ce{O2} that may react with atomic oxygen to form \ce{^{18}O3}. The amount of \ce{^{18}O3} produced in this way is calculated using the band strength determined by \citet{Ioppolo:2013}.

We stress explicitly that even a small efficiency of the studied reaction \ce{H2 + O} may have a substantial impact on water formation for the timescales relevant in space. The nature of the studied system (low reaction probability as well as the low oxygen flux) requires several control experiments to identify the contribution of background water deposition from the different parts of the experimental setup. Therefore, special care has to be taken to exclude any experimental contaminations. To ensure that the amount of background water deposition is as equal as possible on a day to day basis, all (control) experiments are preceded by a day during which the experimental setup has been used only running the \ce{^{18}O2} plasma for 3 hours allowing the fragments to enter also the main chamber in order to obtain stable experimental conditions. Furthermore, the timing of the sequential experimental actions has been kept equal throughout all experiments.

\subsection{Results and Discussion}\label{expRD}
This section explains the principle behind the ten experiments mentioned in Table~\ref{allexps}. Subsequently, the RAIRS results and QMS data are discussed as well as several ways to establish an upper limit of water production. We will show that with the set of performed experiments a conservative upper limit of 0.09~ML is found over an experimental duration of 75 minutes.

To disentangle the origin of the different contributions to the detected 20 amu mass signal in the QMS (experiment 1), three control experiments are performed as indicated in Table~\ref{allexps}: (a) to see the amount of \ce{H2^{18}O} produced inside the plasma (experiment 2), (b) to find the influence of the high \ce{H2} pressure inside the main chamber that can potentially result in sputtering of water off the walls of the UHV system (experiment 3) and (c) to check on the background deposition of water without any atoms or molecules present in the setup (experiment 4). The upper limit to water production is then determined by 
\begin{equation}
 [\ce{H2^{18}O}] \left( {(1)} - {(2)} - {(3)} + {(4)} \right)\;. \label{h2oprod}
\end{equation}
{Experiment 4 is added here, not subtracted. The reason behind this is that experiment 4 gives a contribution that is already included in each other experiment. Therefore if we subtract experiments 2 and 3 from 1, the contribution of experiment 4 is subtracted twice and should therefore be added added once to get the correct number. }
Apart from the aforementioned control experiments, a series of other experiments are performed and added to Table 2 (experiments 5-9). Firstly, we expect the amount of water formed on the sample to be very small. Therefore, we performed experiment 1 for a four times longer duration (experiment 5) to allow for a possible detection of water ice with RAIR spectroscopy. Secondly, we conducted 
{experiments 6, 7 and 8} at different temperatures to retrieve information on the nature of the surface reaction that may lead to the formation of water ice. For instance, the so-called Langmuir-Hinshelwood (LH) mechanism is temperature dependent, whereas the Eley-Rideal (ER) and Hot Atom (HA) mechanisms are much less so. Finally, we performed two more {experiments (9 and 10)} with \ce{D2} instead of \ce{H2} to test to which extent a reaction occurs via (partial) tunneling. Changing the mass of a reactant is a well established experimental technique generally used to verify whether or not a reaction is classically (thermally) activated or proceeds through tunneling {\citep{Oba:2012, Oba:2014}} .

\subsubsection{RAIRS}

In all the experiments where the plasma source was operated, ozone formation was confirmed through RAIRS, but no significant difference could be found between the production in experiments 1 and 2. The amount of \ce{O3} detected in both cases is equal to the total amount of O atoms deposited on the surface within the 30\% uncertainty in the flux. Therefore, this leaves a maximum of 30\% of the O flux to be used for reaction with \ce{H2}, \emph{i.e.}, an upper limit to water production of 
\begin{align*}
 30\% \cdot \frac{ 2\times 10^{11} \text{\small at cm$^{-2}$ s$^{-1}$} \cdot 75 \text{ \small min} \cdot 60 \text{ \small s min$^{-1}$}}{ 1\times 10^{15} \text{ \small at ML$^{-1}$}} 
\end{align*}
amounting to 0.27~ML in 75 minutes.

\begin{figure}[t]
\centering
\includegraphics[width=0.48\textwidth]{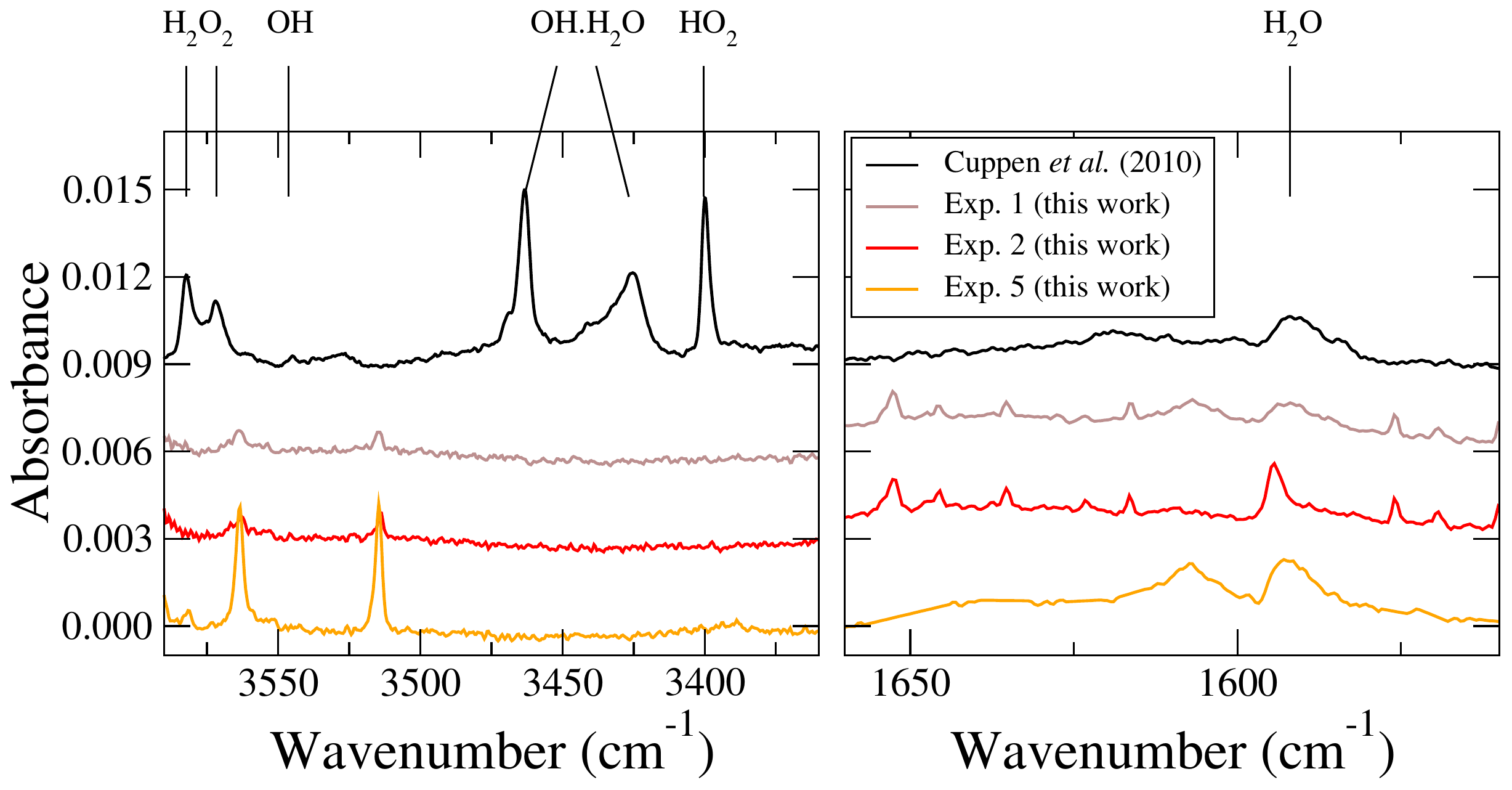}
\caption{RAIR difference spectra from a co-deposition of H and \ce{^{16}O2} from \citet{Cuppen:2010B}, \ce{H2}, \ce{CO2} and $^{18}$\ce{O} (experiment 1), \ce{CO2} and $^{18}$\ce{O} (experiment 2). {Spectra are baseline corrected and offset for clarity. The spectra corresponding with	 experiments 1, 2 and 5 are scaled with a factor 3. Note that the multitude of peaks in the right panel for experiments 1 and 2 are due to water vapor in the setup and the peaks at 3515 and 3564 cm$^{-1}$ are also visible in a `pure' \ce{CO2} spectrum.}}\label{RAIRS}
\end{figure}

Experiment 1 does not result in a detectable amount of formed OH or \ce{H2O} on the basis of their infrared solid state spectral features. Moreover, there is no significant difference between RAIR spectra of experiments 1 and 2, as demonstrated in Fig.~\ref{RAIRS}. The small features visible in the 1600-1800 cm$^{-1}$ range are due to water vapor and they in fact determine the detectable level.
Comparing these spectra to a spectrum obtained from a previous co-deposition experiment of H:\ce{O2} = 1:1 \citep{Cuppen:2010B}, where OH, OH$\cdot$\ce{H2O} and \ce{H2O} spectral bands were found at 3548, 3463, 3426, and 1590 cm$^{-1}$, we conclude that the maximum water production falls below the detection limit of RAIR spectroscopy during a 75 minutes experiment. Therefore, we performed a 300 minutes co-deposition (experiment 5 in Table~\ref{allexps}). In this case, the water peak at 1590 cm$^{-1}$ was clearly visible and, moreover, after gently annealing to 110~K at a ramp of 0.5~K min$^{-1}$ to remove \ce{CO2} and \ce{O2} from the ice, a RAIR spectrum was recorded where approximately 1~ML of water was visible. The upper limit to water production seen with RAIR spectroscopy thus remains $\sim$ 0.25~ML for an experiment of 75 minutes duration.

\subsubsection{QMS}\label{QMS}

\begin{figure}[t]
\centering
\includegraphics[width=0.48\textwidth]{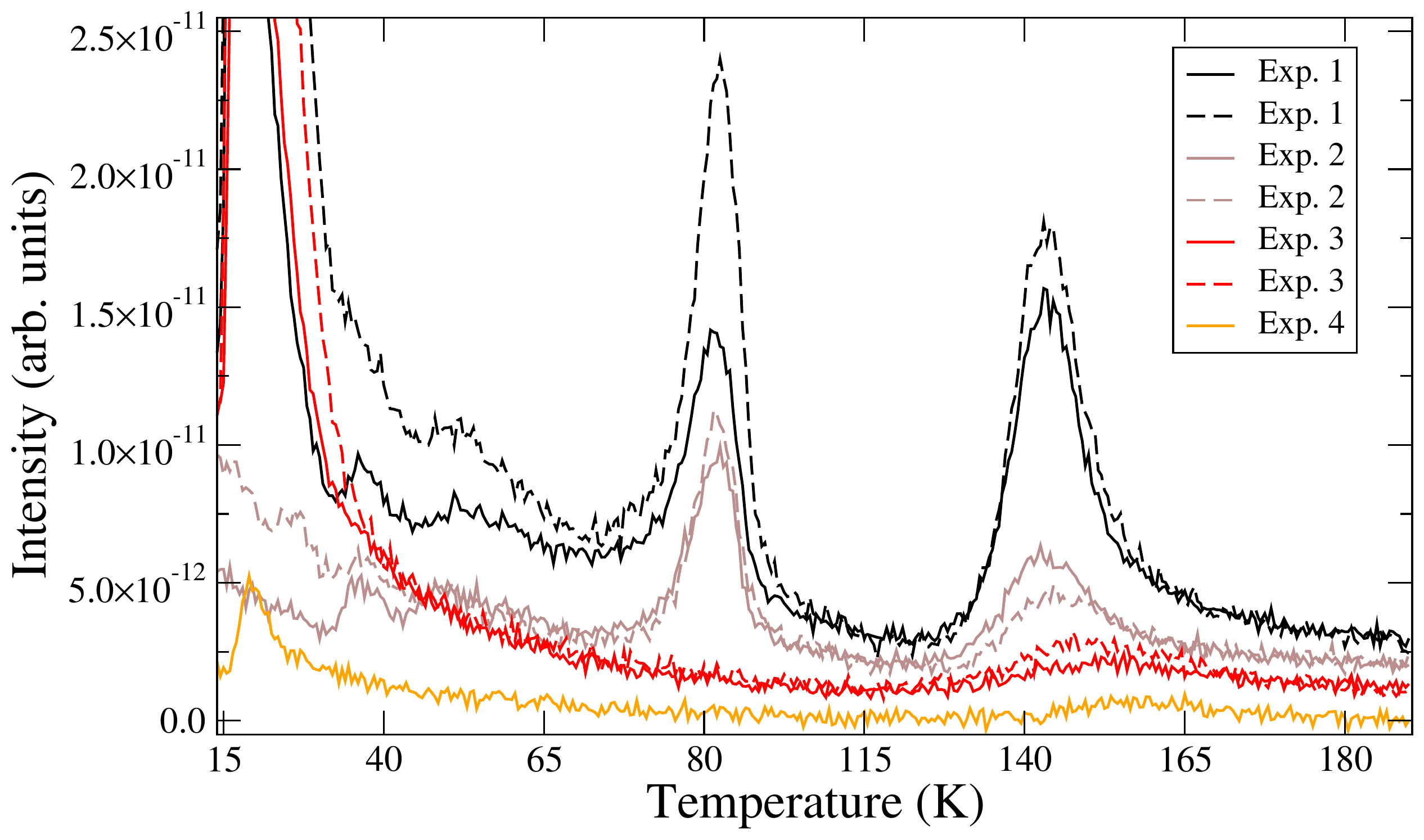}
\caption{QMS traces of mass 20 amu for experiments 1-4 from Table~\ref{allexps}. Spectra are baseline corrected, offset for clarity and binned by averaging 5 points. Experiments 1-3 have been performed twice, hence two traces are depicted by the solid and dashed lines.}\label{QMS20}
\end{figure}

Quadrupole mass spectroscopy allows to better constrain an upper limit for water formation thanks to its larger sensitivity. Table~\ref{allexps} summarizes the integrated baseline corrected QMS signals for mass 20 amu (H$_2^{18}$O). Figure~\ref{QMS20} shows the baseline corrected QMS traces of experiments 1-4 from Table~\ref{allexps}, {both the co-desorption with \ce{CO2} and the thermal desorption of \ce{H2O} are visible}. Experiments 1-3 were performed twice and both traces are shown. The desorption in the region between 14 and 70~K is not taken into further consideration. This is because of the contribution from the species desorbing from the heating tape area in proximity of the substrate as well as because of the oversaturation of the signal by desorption of \ce{H2} or \ce{D2}, as can be concluded from comparing experiment 1 and 3. Experiments 1, 2, and 3 have all been performed twice and the difference between the sum of the integrated signals of two identical experiments is 16, 5, and 26\% respectively, indicating that the overall uncertainty will be of the order of 25\% or smaller. 

The upper limit to water production, calculated with Eqn.~\ref{h2oprod}, is about a factor 2 smaller than concluded from the RAIRS data: 0.14~ML during a 75 minutes experiment. The m/z = 20 signal of both the co-desorption with \ce{CO2} and pure desorption of water is taken into account.

Species that react via the LH mechanism are thermalized and stay on the surface where they diffuse until they meet. This mechanism can be tested by changing the temperature of the ice. In this case, the production of water is expected to decrease with increasing temperature because of a lower surface abundance of \ce{H2} and, moreover, no products should be detected at temperatures above the desorption temperature of one of the reactants. {For this reason, the experimental temperatures employed here are 17, 35 and 50~K (experiments 6, 7 and 8). All detected m/z = 20 signals in these experiments are close to the background level determined at 14~K by experiments 2 and 3. We assume that the observed water is indeed formed - even though this not necessarily has to be the case - and below we discuss various mechanisms. The detected amounts at 17 and 35~K are equal, implying that the LH mechanism is unlikely to be governing any potential reaction, because of the temperature dependence of the residence time at the surface. Moreover, the integrated m/z = 20 signal decreases further when increasing the temperature to 50~K but it still remains non-negligible. } This means that the ER and/or HA mechanisms should be responsible for any \ce{H2O} formation, at least in part and likely even at 14~K. For both mechanisms one or more reaction partners are not thermalized. For the HA mechanism again both reaction partners are present on the surface, but at least one of them is in some excited state (\emph{i.e.}, not thermalized), whereas ER assumes that one reaction partner is present on the surface and the second comes directly from the gas phase and therefore must have a temperature of $\sim$300~K. Both mechanisms in combination with excitation are not expected to be astronomically important, because of the longer time scales and the much lower gas phase temperature in dense molecular clouds. The significance of this reaction pathway in the ISM, therefore, will be negligible.

The reaction itself can proceed either classically activated or through a combination of both a classical and tunneled contribution (\emph{e.g.}, Eqn.~\ref{endotunn}). Tunneling depends on the mass of the reactants involved. Exchanging hydrogen for deuterium would result in a decrease of the tunneled reaction rate of \ce{D2 + O} and therefore a decrease in the production of m/z = 22 (\ce{D2^{18}O}) compared to m/z = 20 (\ce{H2^{18}O}). Comparing the integrated QMS signals of m/z = 22 in experiments {9 and 10} with m/z = 20 in experiment 1 at 125-175~K, we indeed see a large drop up to barely no signal. Therefore, \ce{H2^{18}O} formation in experiment 1 through a mechanism in which tunneling plays a role cannot be eliminated. Because of the endothermicity of the reaction, this has to be a combination of classical and tunneling behavior. As explained above, the classical part can be overcome by some excitation effect.

Finally, even in the experiments performed with \ce{D2} still \ce{H2^{18}O} was detected, which can only be caused by water contamination. From the result found in {experiment 9} it is possible to directly estimate the upper limit with 
\begin{equation}
[\ce{H2^{18}O}] \left( {(1)} - {(9)} \right)
\end{equation}
instead of with Eqn.~\ref{h2oprod}. The difference in signal between {experiments 1 and 9} is therefore taken as the final range for the upper limit to water production for our KMC model, \emph{i.e.} 0.09~ML in 75 minutes. {Because we want to determine an upper limit here, we work with the outcome of experiment 9 and not 10 to guarantee that we remain on the conservative side. }

\section{Theoretical}
\subsection{Kinetic Monte Carlo Model}

This section describes the specific kinetic Monte Carlo procedure used for the simulations and focuses on the difference between modeling experimental results and modeling under interstellar relevant conditions. For a more detailed overview of the method the reader is referred to \citet{Chang:2005} and \citet{Cuppen:ChemRev}. The code used in the present paper is described in \citet{Cuppen:2007A} and \citet{Lamberts:2013, Lamberts:2014}.

The grain is represented by a lattice of $50\times50$ sites with periodic boundary conditions, in which each lattice site can be occupied by one of the following species: H, \ce{H2}, O, \ce{O2}, \ce{O3}, OH, \ce{HO2}, \ce{H2O}, and \ce{H2O2}. Interstitial sites can be occupied only by H, \ce{H2}, O, and OH.
Processes incorporated in the simulations are (i) deposition from the gas phase to the surface, (ii) desorption from the surface back into the gas phase, (iii) diffusion on the surface, (iv) reaction, when two species meet each other, and (v) (photo)dissociation upon energy addition to the species. 
Each of these processes is simply modeled as a change in the occupancy of the sites involved. The event rates are assumed to be classically activated and are calculated using (a form of) Eqn.~\ref{thermal}. The barrier for desorption and diffusion depends on the binding energy of the species to the specific site it occupies. The reaction network consisting of 16 surface reactions and their corresponding rates is taken from \citet{Lamberts:2014}. Photodissociation is implemented only in the interstellar simulations to investigate the influence of the interstellar radiation field. In this case, the five relevant reactions and their rates are taken from \citet{Dishoeck:2006}.

The following strategy is applied: first kinetic Monte Carlo calculations are used to reproduce the experiments with the aim to find an upper limit for the reaction rate (Section~\ref{ExpMod}). The resulting rate is subsequently used to simulate the formation of interstellar ice on astrochemical timescales with a full water surface reaction network to test the contribution of the \ce{H2 + O} reaction to the total production of water ice on interstellar grains in dense clouds (Section~\ref{AstroMod}). Note that, again, this is a conservative method since we already attributed any possible \ce{H2O} formation to mechanisms not relevant in the ISM. Below, however, we will assume a LH type mechanism. Also, our reaction network does not include any species with C or N atoms which will consume hydrogen as well. Here we specifically compare the contributions of the reactions \ce{H + O} and \ce{H2 + O}.

\subsection{Experimental Modeling}\label{ExpMod}
All surface abundances increase linearly with time, similar to those for co-deposition experiments in \citet{Lamberts:2013, Lamberts:2014}. The final abundances mentioned here are after 75 simulated experimental minutes. In all experimental simulations water is produced via the immediate follow-up reaction \ref{R2}, \ce{H + OH -> H2O}, because of our implementation of zero excess energy for the reaction \ce{H2 + O -> OH + H}. H and OH namely remain in each other's vicinity and can thus easily react. The uncertainty in the \ce{H2O} surface abundance is derived by performing two different simulations both three times. We find values decreasing in time from roughly 25 to 7\%, where the largest error bar thus corresponds to the lowest amount of species on the surface.

The values for the fluxes used in the simulations are equal to those listed in Table~\ref{allexps} for the used experiments. The sticking coefficients are assumed unity for the heavier species (\ce{^{18}O}, \ce{^{18}O2} and \ce{CO2}), but is set to a consevative value of 0.2 for \ce{H2}. Experimental results on the sticking of \ce{H2} at 300~K to a 10~K surface indeed indicate such low coefficients \citep{Chaabouni:2012}. The \ce{CO2} flux may be lower due to freeze-out on the cold finger of the cryostat, but, again, to remain on the conservative side we take the highest value of $1.6\times 10^{14}$. The rest of the parameter settings used here are mentioned in Table~\ref{parameters}. There is a certain arbitrariness in the choice which values to use exactly for the input parameters. Here, all input variables are chosen such that they would result in a high reaction rate of the reaction \ce{H2 + O}. This is illustrated by the \ce{H2} sticking coefficient and the flux of \ce{CO2}, namely a low sticking coefficient results in less \ce{H2 + O} encounters and therefore would require a faster reaction rate to produce a result equal to that with a higher coefficient. The same holds for a decrease of the \ce{CO2} flux.

\begin{table}[t]
\centering
\small
\caption{Minimal, maximal and standard parameters used and varied in the experimental simulations.}
\label{parameters}
\begin{tabular}{l@{\;\;}l@{\;\;}lll@{\;\;}l}
\hline
 $E_{\text{diff},\;\ce{H2}}$ & $E_{\text{diff},\;\ce{O}}$ & $T_{\text{surf}}$ & $T_{\text{gas}}$  &  $k_{\ce{O2 + O}}$ & $k_{\ce{H2 + O}}$\\
 (K) & (K) & (K) & (K) & (s$^{-1}$) & (s$^{-1}$) \\
\hline
195 & 330 & 13.5 & 300   & $8.2\times 10^{-5}$ & $1.35\times 10^{-1}$ \\
 250 & 1100 & 13.5 & 300    & $1\times 10^{12}$ & $9.8\times 10^2$ \\
\hline
220 & 1100 & 13.5 & 300 & $8.2\times 10^{-5}$ & $5.1\times 10^1$ / $2.2\times 10^2$  \\
\hline
\end{tabular}
\end{table}

\begin{table}[t]
\centering
\small
\caption{ Summary of the impact of each parameter on the \ce{O3} and \ce{H2O} abundances in the ice. }
\label{summary}
\begin{tabular}{lllllll}
\hline
no. & $k_{\ce{H2 + O}}$ & $k_{\ce{O2 + O}}$ & $E_{\text{diff},\;\ce{H2}}$ & $E_{\text{diff},\;\ce{O}}$ & \ce{H2O} & \ce{O3} \\
 & (s$^{-1}$) & (s$^{-1}$) & (K) & (K) & (ML) & (ML) \\
\hline
1 	&$9.8\times 10^{2}$	& $8.2\times 10^{-5}$ &220 & 1100 &   0.15 & 0.01 \\
2 	&$2.2\times 10^{2}$	& $8.2\times 10^{-5}$ &195 & 1100 &   0.09 &0.01 \\
3 	&$2.2\times 10^{2}$	& $8.2\times 10^{-5}$ &220 & 1100 &   0.09 &0.01 \\
4 	&$2.2\times 10^{2}$	& $1.0\times 10^{12}$ &220 & 1100 &   0.10 &0.05 \\
5 	&$2.2\times 10^{2}$	& $8.2\times 10^{-5}$ &220 & 555  &   0.10 & 0.01\\
6 	&$2.2\times 10^{2}$	& $1.0\times 10^{12}$ &220 & 555  &   0.08 & 0.06\\
7 	&$2.2\times 10^{2}$	& $8.2\times 10^{-5}$ &220 & 330  &   0.07 & 0.01\\
8 	&$2.2\times 10^{2}$	& $8.2\times 10^{-5}$ &225 & 1100 &   0.09 &0.01 \\
9 	&$5.1\times 10^{1}$	& $8.2\times 10^{-5}$ &220 & 1100 &   0.05 &0.01 \\
10 	&$5.5\times 10^{0}$	&$8.2\times 10^{-5}$ & 195 & 1100 &   0.02 &0.01 \\
11 	&$5.5\times 10^{0}$	&$8.2\times 10^{-5}$ & 220 & 1100 &   0.03 &0.01 \\
12	&$5.5\times 10^{0}$	& $8.2\times 10^{-5}$ &220 & 555  &   0.03 & 0.01\\
13 	&$5.5\times 10^{0}$	&$8.2\times 10^{-5}$ & 220 & 330  &   0.02 & 0.00\\
14 	&$5.5\times 10^{0}$	&$8.2\times 10^{-5}$ & 225 & 1100 &   0.03 &0.01 \\
15 	&$1.35\times 10^{-1}$ 	& $8.2\times 10^{-5}$ & 220 & 1100 &   0.00 & 0.02 \\
\hline
\end{tabular}
\tablefoot{ Abundances given here are scaled to 75 minutes where appropriate.}
\end{table}

{The approach taken here is to find a set of parameters that allows to reproduce the experimental upper limit of 0.09~ML (see Section~\ref{expRD}) in 75 minutes of experiment. In order to do so, we varied several parameters, as mentioned in Table~\ref{parameters}. Firstly, the diffusion barrier of \ce{H2} is set to 195, 220 and 250~K. Next, we performed simulations using barriers for oxygen atom diffusion with 330, 555, and 1100~K. The latter value has been used in earlier studies \citep{Lamberts:2013, Lamberts:2014} and the second value is half of this number. Very recently, literature values have become available (\emph{e.g.}, \citet{Lee:2014, Congiu:2014}) that predict values between 350 and 1000~K, the domain embedded by our chosen barrier values. The reaction rates of the reactions \ce{O2 + O} and \ce{H2 + O} have also been varied. The first reaction rate was set to the value used in a previous study (\citep{Lamberts:2013}, $8.2\times 10^{-5}$ s$^{-1}$) as well as a value corresponding to a barrierless reaction ($1.0\times 10^{12}$ s$^{-1}$). The second rate has been set to $1.35\times 10^{-1}$, $5.5$, $5.1\times 10^1$, $2.2\times 10^2$, and $9.8\times 10^2$ s$^{-1}$. These values represent exactly the range in which the reaction \ce{H2 + O} becomes effective in competing with diffusion and other reactions. In other words, for reaction rates below  $1.35\times 10^{-1}$ s$^{-1}$ the reaction does not occur at all. This sensitive window of reaction rates has been found by performing several test simulations used to probe the influence of the parameters. 
We started with two models for each parameter, using the minimal and maximal value while keeping all other parameters constant to their standard value as indicated in the final row of Table~\ref{parameters}. Due to the dominant role of $k_{\ce{H2 + O}}$, the influence of any other parameter was typically checked at two different reaction rates.
Only in the case that a dependence on a particular parameter was found, we varied that specific parameter to other values in additional simulations while keeping other parameters constant to their standard value. Therefore, not a full grid of models was used, but rather a total of 15 simulations have been performed. The resulting \ce{O3} and \ce{H2O} abundances are summarized in Table~\ref{summary}.}

{The diffusion rates of both O and \ce{H2} only play a role when the reaction with the O atom is almost prohibited. In this case, a high diffusion rate leads to a lower water production, because of the favorable competition with respect to reaction. The amount of \ce{O3} produced in the simulations does not depend on the diffusion rate of oxygen atoms, but does show a strong dependence on the reaction rate of \ce{O2 + O}. Previously we used a reaction rate of $8.2\times 10^{-5}$ s$^{-1}$ \citep{Lamberts:2013}. Here we see that a faster rate is needed to reproduce the  amounts detected by RAIR spectroscopy. We will come back to this in the next section.} 

\begin{figure}[t]
\centering
\includegraphics[width=0.48\textwidth]{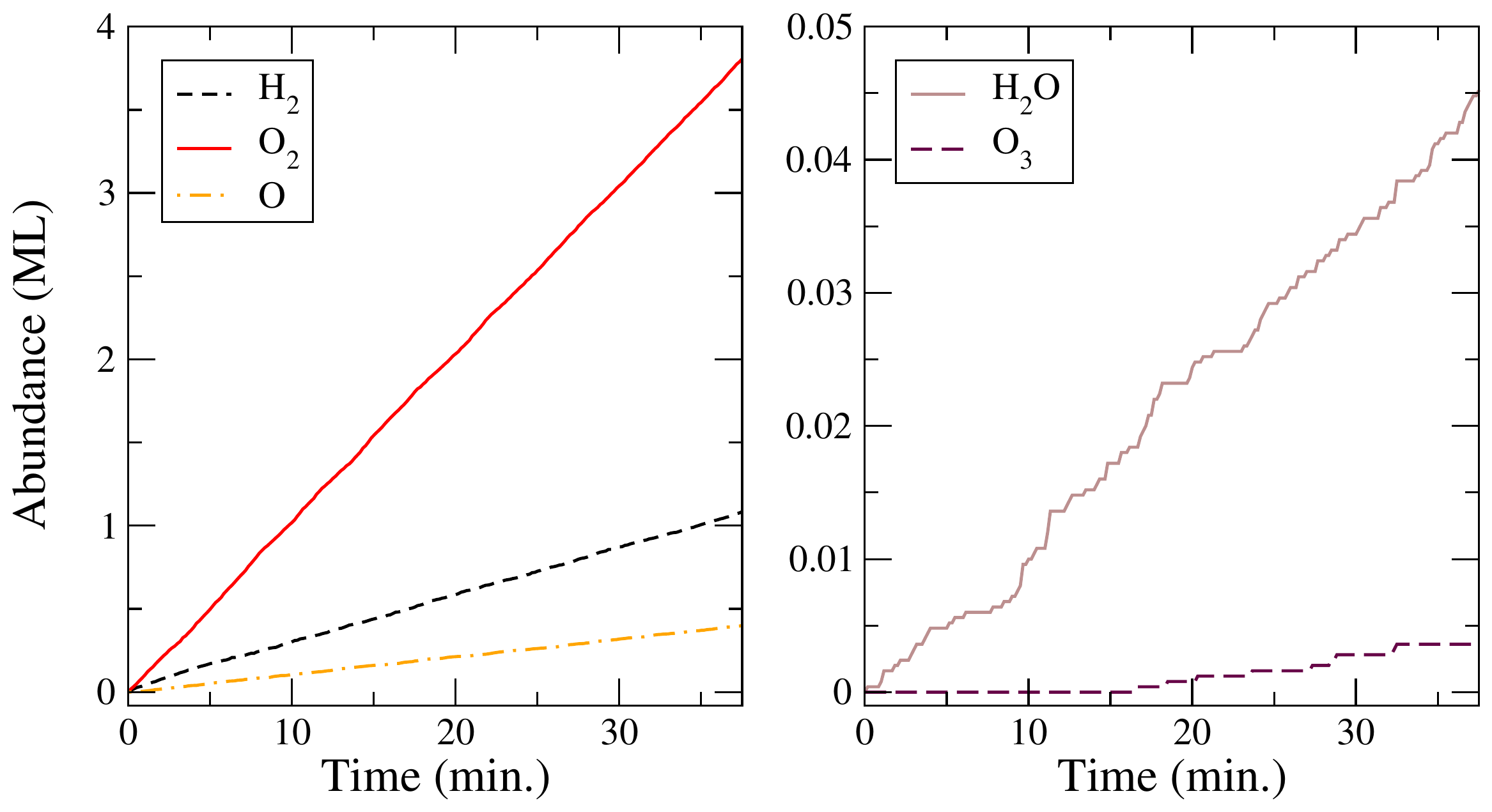}
\caption{{Surface abundances of O, \ce{O2}, \ce{O3}, \ce{H2}, and \ce{H2O} in time for the `upper limit' simulation. One should realize that the total amount of deposited ice over the course of this simulation is 360 ML. The dominant component, by far, is \ce{CO2} (not shown) because of its high flux. }}\label{model}
\end{figure}

{From Table~\ref{summary} it can be deduced that for the production of water the parameter that has the largest impact on the final abundances is the reaction rate itself, namely $k_{\ce{H2 + O}}$. For most simulations the final \ce{H2O} abundance remains below the experimental upper limit of 0.09~ML.
In Fig.~\ref{model}, the surface abundances of O, \ce{O2}, \ce{O3}, \ce{H2} and \ce{H2O} are depicted over a simulated period of 37.5 minutes for simulation number 3 in Table~\ref{summary}, which we define as the  `upper limit' simulation. It concerns a co-deposition experiment, therefore the profile of surface abundances is increasing linearly with time. The high amount of \ce{H2} should be interpreted as 1.1~ML distributed over the total ice thickness of 360~ML. The total ice thickness is mainly determined by the high \ce{CO2} flux and therefore, a deposited \ce{H2} molecule either reacts, desorbs or is covered by another \ce{CO2} molecule. That means that on average there is 0.003~ML of \ce{H2} in each monolayer and thus this corresponds to the average surface coverage at any given time. The final \ce{H2O} abundance in this figure is 0.045~ML, because of the reduced time scale. The value of $k_{\text{exp. max}}(13.5~\text{K}) = 2.2\times 10^2$ s$^{-1}$ leads to this \ce{H2O} production that corresponds to the experimentally determined value. This rate will be used to simulate water formation in the ISM through \ce{H2 + O}. }

\subsection{Astrochemical Modeling}\label{AstroMod}
Two dense clouds with different temperature, density and UV field are studied here. Their physical parameters are chosen identical to those of dense clouds I and II in \citet{Lamberts:2014} as summarized in Table~\ref{ismsims}. The high densities $n_\text{H}$ and simultaneous low temperatures, but high $A_\text{V}$ values mimic typical values found in dense clouds. A major difference between the present and previous work is the inclusion of endothermicity of reaction~\ref{R1}. In the preceding study, we included an excess energy of 1400~K for each reaction in the water formation network with two reaction products and the energy is spread over these products. The excess energy for the endothermic reaction \ce{H2 + O} is therefore explicitly set to 0~K, all other two-product reactions obtain a reaction heat of 1400~K. {We use the same full water reaction network, but following the outcome of \citet{Lamberts:2013, Lamberts:2014}, we omitted the reaction channel \ce{H + HO2 -> H2O2}. The network used here consists of a total of 16 reactions. }

The main parameter varied in the astrochemical simulations, is the rate of reaction \ref{R1} ranging between the fastest $k_{\text{exo. tunn.}}$ and the slowest $k_{\text{endo. tunn.}}$ as explained in Section~\ref{CalculateRate}. From the experiments it is deduced in Section~\ref{expRD} that \emph{if} water is produced starting from \ce{H2 + O} it can be only via a mechanism involving overcoming the endothermicity classically `followed' by tunneling through the barrier as indicated in Fig.~\ref{ReactionPathway} and Eqn.~\ref{endotunn}. Scaling the reaction rate determined experimentally at 13.5~K to rates relevant at 10 and 12~K - the surface temperatures of the grains in the dense cloud studied here - is realized through the approach outlined below:  
\begin{align}
k_{\text{exp. max}}(13.5~\text{K}) &= C \cdot \exp\left(- \frac{\Delta E}{T} \right) \nonumber \\
 2.2\times 10^2 &=  C \cdot \exp\left(- \frac{960}{13.5} \right) \nonumber\\
&\Rightarrow C = 1.68\times10^{33} \nonumber \\
k_{\text{exp. max}}(T) &=  1.68\times10^{33} \cdot \exp\left(- \frac{960}{T} \right) \label{upperlimit}\;.
\end{align}
Here, we assume that the endothermicity of the reaction, $\Delta E$, is well constrained by the gas-phase value of 960~K. The tunneling mechanism, activation energy and pre-exponential factor are not specifically considered (compare to Eqn.~\ref{endotunn}), but are all combined in the factor \emph{C}, which is considered temperature independent over the small temperature range studied here.

\begin{table}[b]
\centering
\small
\caption{Parameters used in the astrochemical simulations, \emph{i.e.}, dense clouds I and II (1). }
\label{ismsims}
\begin{tabular}{lllllll}
\hline
 & $A_\text{V}$ &  $n_\text{H}$ & $n_\text{H(I)}$ & $n_\text{O(I)}$ & $T_\text{gas}$  &$T_\text{grain}$  \\
 & 		& (cm$^{-3}$) & (cm$^{-3}$)  & (cm$^{-3}$) & (K) &(K) \\
\hline
I & 5 & $5\times 10^3$ & 2 & 1.5 & 20 & 12 \\
II & 10 & $2\times 10^4$ & 2 & 6 & 10 & 10 \\
\hline
\end{tabular}
\tablebib{(1) \citet{Lamberts:2014}}
\end{table}

\begin{table*}[t]
\small
 \centering
\caption{Contributions of the different surface reaction routes to OH and \ce{H2O} formation after a coverage of 1~ML is reached and total produced water rate for dense clouds I and II for different values of $k_{\ce{H2 + O}}$.}\label{contributionsOH}
\begin{tabular}{l@{\;\;}l@{\;\;}l|c@{ }c@{ }c@{ }c@{ }c|c@{\;\;}c|c}
\hline
\multicolumn{1}{l}{Cloud} &     	& 	&    \multicolumn{1}{l}{\ce{H2 + O} }   &  \multicolumn{1}{l}{\ce{H + O} } &  \multicolumn{1}{l}{\ce{H + HO2} } &  \multicolumn{1}{l}{\ce{H + O3} } 	&  \multicolumn{1}{l|}{\ce{H + H2O2}}   &  \multicolumn{1}{l}{\ce{H + OH}} & \multicolumn{1}{l|}{\ce{H2 + OH}} &  \ce{H2O} prod.  \\

			  & 		& 	& \multicolumn{1}{r}{\ce{\rightarrow \ce{OH + H}}}& \multicolumn{1}{r}{\ce{\rightarrow \ce{OH}}} & \multicolumn{1}{r}{\ce{\rightarrow \ce{OH + OH}}} & \multicolumn{1}{r}{\ce{\rightarrow \ce{OH + O2}}} & \multicolumn{1}{r|}{\ce{\rightarrow \ce{OH + O}}} & \multicolumn{1}{r}{\ce{\rightarrow \ce{H2O}}} & \multicolumn{1}{r|}{\ce{\rightarrow \ce{H2O + H}}} & \\ 

	  &        $k_{\ce{H2 + O}}$       	&  \;\;\;{(s$^{-1})$}	&  	(\%)		&  	(\%)	& 	 (\%)	&  (\%)	&  (\%)	&  (\%)	&  (\%)	&  (ML kyr$^{-1}$) \\
\hline
I         & $k_{\text{exo. tunn.}}$$^{a}$   	&  = $2.3\times 10^{5}$		&  95.0	&  4.5	&$\sim$0&$\sim$0&$\sim$0& $\sim$100 	& $\sim$0 & 0.19	\\
I         & $k_{\text{exp. max.}}$$^{b}$ 	&  = $3.0\times 10^{-2}$	& \textbf{11.0}	&  74.0	& 12.0 	&  0.5 	& 2.5 	& 86.5		& 9.0	& 0.15  \\
I         & $k_{\text{endo. tunn}}$$^{c}$  	&  = $4.4\times 10^{-29}$	&  0	&  82.0	& 14.0	& 1.0  	& 3.0 	& 75.5		& 19.5 	& 0.14 \\
II        & $k_{\text{exo. tunn.}}$$^{a}$ 	&  = $2.3\times 10^{5}$		&  98.5	&  1.5 	&$\sim$0& 0	&$\sim$0& 99.0 		& 1.0 	& 0.54 \\
II        & $k_{\text{exp. max.}}$$^{b}$ 	&  = $3.5\times 10^{-9}$	&  \textbf{1.0}	&  56.0	& 31.0 	& 6.0	& 6.0 	& 26.0		& 60.0	& 0.24   \\
II        & $k_{\text{endo. tunn}}$$^{c}$  	&  = $5.0\times 10^{-36}$	&  0	&  54.0	& 35.0 	& 5.5	& 5.5	& 20.5		& 63.5	& 0.24 \\
\hline                                                               
\end{tabular}
\tablefoot{
\tablefoottext{a}{Eqn.~\ref{tunneled}, with $E_{\text{a}}$ equal to $E_{\text{a}} + \Delta E$ from Fig.~\ref{ReactionPathway}, \emph{i.e.} 3000~K. } 
\tablefoottext{b}{Eqn.~\ref{upperlimit}.} 
\tablefoottext{c}{Eqn.~\ref{endotunn}, with $E_{\text{a}}$ = 2040~K and $\Delta E$ = 960~K.} 
}
\end{table*}

Table~\ref{contributionsOH} gives the contributions of the different surface reaction routes to \ce{OH} and \ce{H2O} formation and the total amount of \ce{H2O} produced per kyr in the simulations. Three different reaction rates are considered: (i) assuming exothermic tunneling with Eqn.~\ref{tunneled}, (ii) using the experimentally determined maximum rate with Eqn.~\ref{upperlimit}, and (iii) assuming that $E_\text{a} + \Delta E = $ 3000~K in Eqn.~\ref{endotunn}. The results presented here are obtained at a time of $\sim2.0\times 10^{4}$ and $\sim3.5\times 10^3$ years for the two clouds respectively. This may seem too short to be interstellar relevant, and is due to the high computational costs, but in fact all abundances increase linearly or reach a steady state abundance before this time. Moreove, all values are calculated after the grain has already been covered with a total of 1~ML of species.

The following reaction channels are considered in Table~\ref{contributionsOH}. Firstly, the production of the OH radical is broken down into the separate contributions of five reaction routes, namely \ce{H2 + O}, \ce{H + O}, \ce{H + HO2}, \ce{H + O3}, and \ce{H + H2O2}. In the case of cloud I, changing the reaction rate of \ref{R1} simply shifts the main production route from \ce{H2 + O} to \ce{H + O} for decreasing rates. For cloud II, however, there is more oxygen than atomic H present in the cloud. Allowing the reaction \ce{H2 + O} to proceed thus leads to a much higher OH production.

The formation of water can also proceed via multiple reaction routes, but the two that are important here are \ce{H + OH} and \ce{H2 + OH}. In the lower density case, the total water production rate does not change substantially between the three rates. At higher densities, the larger abundance of OH translates immediately into a larger amount of produced \ce{H2O}, since the products of reaction~\ref{R1}, \emph{i.e.}, H and OH, remain again in each other's vicinity. 

Furthermore, focusing on the upper limit to the reaction rate, $k_\text{exp. max.}$, Table~\ref{contributionsOH} clarifies that the reaction \ce{H2 + O} only contributes to a maximum of 11\% to the formation of OH on the surface of dust grains in cloud I and does not contribute at all in cloud II. Since we chose all parameters in such a way to stay on the conservative side, this is an absolute upper limit. Higher \ce{H2} sticking probabilities, lower \ce{CO2} flux due to freeze out on the cold finger or non-thermalized effects as detailed in Sections~\ref{expRD} and~\ref{ExpMod} all lead to lower rates.

{The effect of the O diffusion barrier has been investigated by simulating with the values 555 and 1100~K}. Although the total water production does not change much, the relative contributions of the reaction producing OH radicals do: with a faster \ce{O} diffusion, the competition between diffusion and the reaction \ce{H2 + O} favors diffusion, leaving \ce{O} free on the surface to react with other species. Consequently, the reactions \ce{H + O}, \ce{O + O} or \ce{O + O2} play a larger role, the extent of which depends on the density. Furthermore, increasing the reaction rate for \ce{O3} formation results in a larger contribution of the reaction channel \ce{H + O3}. In the end, also these effects will decrease the efficiency of the reaction \ce{H2 + O}.

\section{Astrophysical Implications}
Since the reaction \ce{H2 + O} only contributes to a maximum of 11\% to the formation of OH, water formation is dominated by the other reaction routes, such as \ce{O + H}, \ce{O2 + H}, \ce{OH + H} and \ce{OH + H2}. This implies that depending on the ratio of O/H in the gas phase, the limiting factor to the water formation rate in dark clouds is the amount of H atoms available. Additionally, for high O/H ratios, a higher diffusion rate of O atoms can lead to more reactions of the type \ce{O + O} \citep{Congiu:2014}. This does not mean that water formation is prohibited, since the reaction channel \ce{O2 + H} can also lead to efficient water formation \citep{Ioppolo:2010, Cuppen:2010B, Lamberts:2013}.

The experimentally found upper limit for the reaction rate, Eqn.~\ref{upperlimit}, can be compared to the values of the reaction rates where exothermic tunneling was assumed. Taking the final two entries of Table~\ref{rates}, we see that these rates (at 10~K) are always larger. Therefore, the assumed importance of the reaction \ce{H2 + O} for the deuterium fractionation ratios of water on the surfaces of dust grains has to be considered with care \citep{Cazaux:2010, Cazaux:2011}. Their HDO/\ce{H2O} ratio found at low temperatures results from the assumption that the reaction  \ce{HD + O} proceeds via tunneling and therefore mainly produces OH + D. The amount of HDO formed on the surface of dust might in fact be much larger, depending on the main water formation route in the specific region in the interstellar medium (through atomic or molecular oxygen). 

\section{Conclusions}
We presented a combined experimental and modeling study dedicated to study the significance of the reaction \ce{H2 + O -> H + OH} in the framework of the solid state water formation network in interstellar ice (analogues).

From precisely executed Temperature Programmed Desorption experiments in an UHV setup bringing together \ce{H2} and \ce{O} in a matrix of \ce{CO2}, we established an experimental upper limit of the water production. In case this amount of water is indeed being produced on the surface, instead of coming from an additional source of contamination, we find that this can only be caused by a combined classical and tunneled reaction mechanism, based on Eqn.~\ref{endotunn}. An upper limit for the reaction rate was found using a microscopic kinetic Monte Carlo model that converts the maximum number of molecules formed into a possible reaction rate: $1.68\times10^{33} \cdot \exp\left(- {960}/{T} \right)$ s$^{-1}$. Incorporating this rate into simulations run under astrochemically relevant parameters, we find that the reaction \ce{H2 + O} does not contribute to more than 11\% to the formation of water in dense clouds in the interstellar medium. 

This number is an absolute upper limit, as all numbers used are conservative estimates. It is likely that in space the efficiency will be substantially lower.

\begin{acknowledgements}
H.M.C. is grateful for support from the VIDI research program 700.10.427, which is financed by The Netherlands Organisation for Scientific Research (NWO) and from the European Research Council (ERC-2010-StG, Grant Agreement no. 259510-KISMOL). T.L. is supported by the Dutch Astrochemistry Network financed by The Netherlands Organisation for Scientific Research (NWO). Support for S.I. from the Niels Stensen Fellowship and the Marie Curie Fellowship (FP7-PEOPLE-2011-IOF-300957) is gratefully acknowledged. The SLA group has received funding from the European Community’s Seventh Framework Programme (FP7/2007- 2013) under grant agreement no. 238258, the Netherlands Research School for Astronomy (NOVA) and from the Netherlands Organization for Scientific Research (NWO) through a VICI grant. 
\end{acknowledgements}

\bibliographystyle{aa}

\end{document}